\documentclass[aps,prd,preprintnumbers,groupedaddress,showpacs,nofootinbib,floatfix]{revtex4}
\usepackage{amsmath,amssymb,amsbsy}
\usepackage{graphicx,subfigure,psfrag}

\begin{document}

\preprint{UW-PT 04-06}

\title{The phase diagram of twisted mass lattice QCD}

%
\author{Stephen R. Sharpe}
\email[]{sharpe@phys.washington.edu}
\author{Jackson M. S. Wu}
\email[]{jmw@phys.washington.edu}
\affiliation{Physics Department, Box 351560, University of Washington,
Seattle, Washington 98195}

\date{\today}

\begin{abstract}
We use the effective chiral Lagrangian
to analyze the phase diagram of two-flavor twisted mass lattice QCD
as a function of the normal and twisted masses, generalizing
previous work for the untwisted theory.
We first determine the chiral Lagrangian including discretization
effects up to next-to-leading order (NLO) in a combined expansion in which
$m_\pi^2/(4\pi f_\pi)^2 \sim a \Lambda$ ($a$ being the lattice
spacing, and $\Lambda = \Lambda_{\rm QCD}$).
We then focus on the region where 
$m_\pi^2/(4\pi f_\pi)^2 \sim (a \Lambda)^2$,
in which case competition between leading and NLO terms
can lead to phase transitions.
As for untwisted Wilson fermions, we find two possible
phase diagrams, depending on the sign of a coefficient in the chiral Lagrangian.
For one sign, there is an Aoki phase for pure Wilson fermions,
with flavor and parity broken, but this is washed out into a crossover
if the twisted mass is non-vanishing.
For the other sign, there is a first order transition for pure Wilson fermions,
and we find that this transition extends into the twisted mass plane,
ending with two symmetrical
second order points at which the mass of the neutral pion vanishes.
We provide graphs of the condensate and pion masses for both scenarios,
and note a simple mathematical relation between them.
These results may be of importance to numerical simulations.
\end{abstract}

\pacs{11.30.Hv, 11.30.Rd, 12.39.Fe, 12.38.Gc}

\maketitle

\section{\label{sec:intro} Introduction and Conclusion}

There has been considerable recent interest in the
twisted mass formulation of lattice QCD (tmLQCD).\footnote{%
For a recent review see Ref.~\cite{FrezLatt04}.}
This formulation has several significant advantages
compared to ``untwisted'' Wilson fermions:
first, the twisted mass provides an infra-red cut-off for small eigenvalues
of the lattice Dirac operator and thus avoids so-called
exceptional configurations~\cite{Frez00,Frez01};
second, for maximal twisting (i.e. a purely twisted mass term) physical quantities
are automatically O(a) improved~\cite{FR2003};
and, finally, calculations of weak matrix elements are considerably
simplified~\cite{Frez01,Pena04,Frez04}.
It may thus serve, for practical simulations, as an intermediate formulation
between improved staggered fermions (which share the above advantages, but
have the disadvantage of taking the square- or fourth-root of the
fermion determinant) and chiral lattice fermions (which are more computationally
expensive).

We consider here simplest version of tmLQCD which, by construction,
is equivalent in the continuum limit to QCD with two degenerate flavors.
At non-zero lattice spacing, however,
 the flavor group is explicitly broken from $SU(2)$
down to $U(1)$. This flavor breaking is analogous to the ``taste symmetry'' breaking
of staggered fermions that plays a major role in practical simulations.
It is thus important to study the impact of flavor breaking in simulations
of tmLQCD. This can be done analytically using the chiral effective theory
including the effects of discretization. The methodology for doing so for
Wilson fermions was worked out in Ref.~\cite{SS98}.
Here we provide the generalization to tmLQCD, determining the 
chiral effective Lagrangian
to next-to-leading order (NLO).
This extends the work of Ref.~\cite{MS04}
by including terms proportional to $a^2$ ($a$ being the lattice spacing).

There are many possible applications of the resulting effective Lagrangian.
In this paper we focus on the phase structure.
We extend the analysis of Ref.~\cite{SS98} (and the similar considerations
of Ref.~\cite{Creutz}) into the twisted mass plane.
For untwisted Wilson fermions, Ref.~\cite{SS98} found two possibilities as the
(untwisted) quark mass approaches the critical mass.
In the first, the pion masses vanishes, and the theory enters an Aoki phase,
in which flavor and parity are spontaneously broken.
This is the scenario proposed long ago by Aoki~\cite{Aoki84,Aoki86a,Aoki86b},
and supported by results from quenched simulations. 
The phase has width $\Delta m \sim a^2$,
in terms of the physical quark mass $m$.
The second possibility is that there is a first order transition
at which the condensate flips sign, and the pion mass reaches
a minimum, but is non-vanishing. In this scenario
flavor and parity are unbroken.
The choice of scenario is determined by the sign of a particular coefficient
in the chiral Lagrangian [the coefficient is $c_2$ defined in
eqs.~(\ref{E:PE},\ref{E:ChLcoefs})], the value of which depends on the gauge action
and the coupling constant. 
Indeed, numerical work with unquenched Wilson fermions~\cite{Ilgenfritz03}
finds evidence for an Aoki phase at strong coupling (consistent with
the expectations of analytic results in the strong coupling,
large $N_c$ limit~\cite{Aoki84}),
but also finds that the phase disappears as the coupling is weakened.
This suggests that the second scenario applies for moderate and weak couplings,
at least with the Wilson gauge action.
This conclusion is supported by the very recent work of Ref.~\cite{Farchioni04},
who find clear evidence of a first order transition along the Wilson axis.

\begin{figure}
\centering
\subfigure[Phase diagram for $c_2 > 0$]{
\psfrag{b}{\large $\beta$}
\psfrag{a}{\large $\alpha$}
\label{fig:mmu:a}
\includegraphics[width=2in]{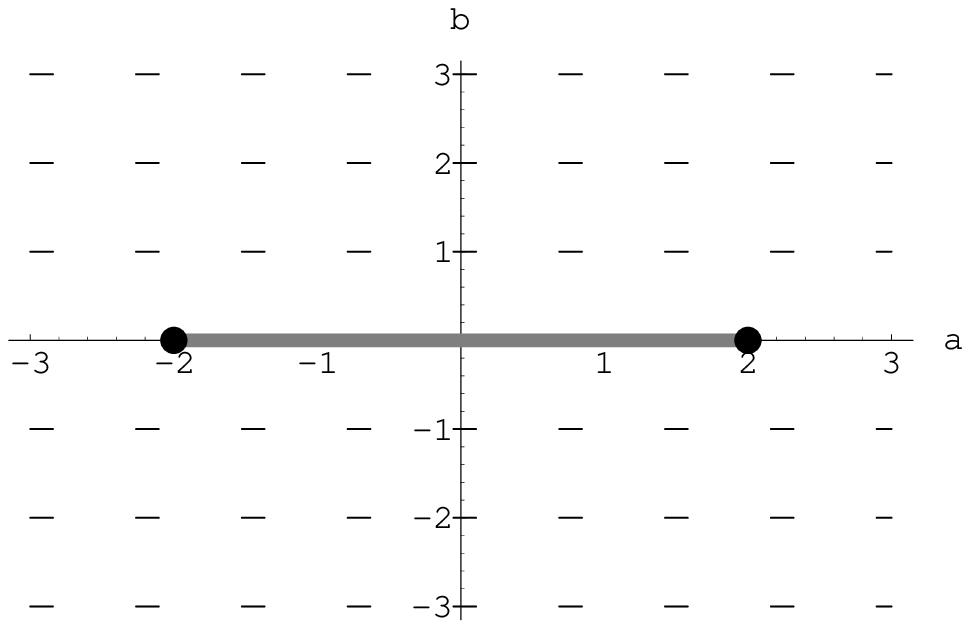}}
\hspace{0.2in}
\subfigure[Phase diagram for $c_2 < 0$]{
\psfrag{b}{\large $\beta$}
\psfrag{a}{\large $\alpha$}
\label{fig:mmu:b}
\includegraphics[width=2in]{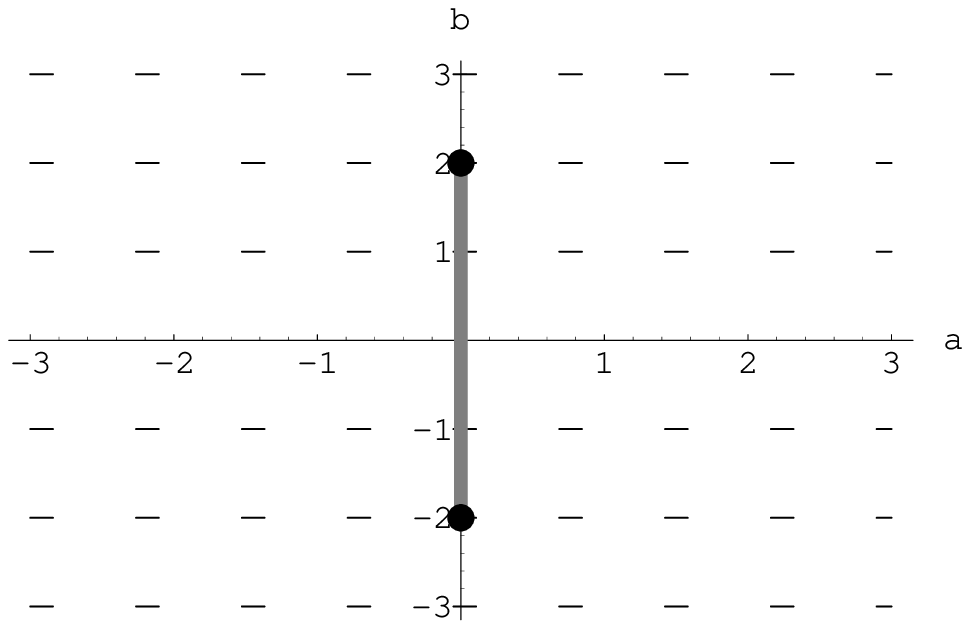}}
\caption{\label{fig:mmu} Phase diagram of tmLQCD. $\alpha$ and $\beta$
  are proportional to the untwisted and twisted mass, respectively,
 in units proportional
  to the lattice spacing squared [see eqs.~(\ref{eq:alphadef},\ref{eq:betadef})]. 
The sign of the coefficient $c_2$
  determines whether flavor symmetry is spontaneously broken in the
  standard Wilson theory.  The solid lines are first order
transitions across which the condensate is discontinuous, with
second-order endpoints. Figures~\ref{fig:c2gt0Am}-\ref{fig:c2lt0}
 show the dependence of
the condensate and pion masses along the horizontal dashed lines.
See text for further discussion.} 
\end{figure}

The extension of the analysis of Ref.~\cite{SS98} into the twisted mass
plane turns out to be relatively straightforward.
The twisted mass introduces only a single additional term in the potential,
one that attempts to twist the condensate in the direction of the full mass.
We find that any non-zero value for the twisted mass washes out the
Aoki phase of the first scenario, as expected when the would-be spontaneously
broken symmetry is explicitly broken.
Thus the Aoki phase itself is confined to a short segment of the untwisted
axis, as shown in Fig.~\ref{fig:mmu}(a). Traversing this segment in
the twisted direction there is a first order transition, with a
discontinuity in the chiral condensate. More interesting, and, in
light of the results of simulations 
noted above, probably more relevant, is the impact of twisting
on the scenario with a first order transition.
Here we find that the transition extends a distance of size $\Delta \mu \sim a^2$
(where $\mu$ is the physical twisted mass) into the plane, as shown 
in Fig.~\ref{fig:mmu}(b).
The transition weakens as $|\mu|$ increases, ending with second-order
points where the neutral pion mass vanishes.

Looking at the phase diagram, the two scenarios appear related
by a $90^\circ$ rotation. In fact, this relation is exact 
for the condensate and the neutral pion mass, 
aside from corrections of $O(a^3)$ in the chiral effective theory.
In other words, approaching the critical mass along the twisted mass axis
the neutral pion mass has exactly the same dependence on
the twisted mass as it does on the normal mass in an untwisted Wilson simulation
{\em with the opposite value of $c_2$}. Similarly, the condensates in the two cases
are simply related by a $90^\circ$ twist. This identity does not, however,
hold for the charged pion masses. These vanish in the Aoki phase, but do not
vanish along the first order transition line for $c_2 < 0$.
Nevertheless, there is still a simple relation between the two scenarios:
the charged pion mass-squareds for $c_2 <0$ are obtained from those
for $c_2>0$ by rotating by $90^\circ$ and then adding
a constant positive offset (proportional to $|c_2| a^2$).

The possible presence of  phase structure extending in the
twisted mass direction may hinder numerical simulations---there would
be premature critical slowing down due to the second-order endpoint,
and metastability beyond~\cite{Farchioni04}. 
One can reduce these effects either by
working at weaker coupling (since $\Delta\mu \propto a^2$), or
by attempting to find a gauge action which gives rise to a smaller
value of $c_2$.

The rest of this paper is organized as follows. In the next section, we present the
two-step process of obtaining an effective chiral theory 
 that describes the long distance physics of the underlying
lattice theory of tmLQCD including discretization effects. 
We then, in Sec.~\ref{sec:anaAoki}, explain how
to use the resulting chiral Lagrangian to investigate the phase diagram
of tmLQCD.

An partial account of the work presented here was given
in Ref.~\cite{Wulat04}. Results on the masses and decay constants
of pions will be presented elsewhere~\cite{SWPrep}.

Similar conclusions have been reached
independently by M\"unster~\cite{Munsterlat04,Munster04}
and Scorzato~\cite{Scorzato}.


\section{\label{sec:ECL} Effective chiral theory}

The theory we consider here is tmLQCD with a
doublet of degenerate quarks.\footnote{%
We do not consider the generalization to non-degenerate
quarks discussed in Ref.~\cite{tmnondegen}.}
In this section, we start by
briefly reviewing the symmetry properties of the tmLQCD action. We
obtain the effective chiral theory for pions by first determining
the effective continuum Lagrangian at the quark level and then matching it 
onto the chiral Lagrangian.

\subsection{Twisted mass lattice QCD}

The fermionic part of the Euclidean lattice action has the
form~\cite{Frez01,FR2003}
\begin{align} \label{E:action}
S^L_F =& \; a^4 \sum_{x} \bar{\psi_l}(x)
\Big[\frac{1}{2} \sum_{\mu} \gamma_\mu (\nabla^\star_\mu + \nabla_\mu)
- r \frac{a}{2} \sum_{\mu} \nabla^\star_\mu \nabla_\mu 
%
%
+ m_0 + i \gamma_5 \tau_3 \mu_0
\Big] \psi_l(x),
\end{align}
where $\psi_l$ and $\bar\psi_l$ are the bare lattice fields
(with ``$l$" standing for lattice and not indicating left-handed),
and $\nabla_\mu$ and $\nabla^\star_\mu$ are the usual covariant
 forward and backward lattice derivatives, respectively.
The bare normal mass, $m_0$, and the bare twisted mass, $\mu_0$, are
taken to be proportional to the identity matrix in flavor space.  A
possible ``twist'' in the Wilson term has not been included because it
can be rotated away by appropriate changes in $m_0$ and $\mu_0$.  This
means that we are using the so-called ``twisted basis'', which we
find simplifies the subsequent analysis. We will not
need to specify the irrelevant parameter $r$ in our analysis, other
than that it should be a non-vanishing constant, satisfying $|r|\le 1$ for
reflection positivity~\cite{FR2003}. 
What we have in mind is the value $r=1$ used in
most numerical simulations.

The action given in (\ref{E:action}), combined with any hypercubically
invariant gluon action, shares the following symmetries with standard
Wilson fermions: invariance under gauge transformations, lattice
rotations and translations, charge conjugation, and the $U(1)$
transformations associated with fermion number.  
  It differs from standard
Wilson fermions in two important ways.  First, the flavor $SU(2)$
symmetry is broken explicitly by the $\mu_0$ term down to the $U(1)$
subgroup with diagonal generator $\tau_3$.  Second, ordinary parity
$\mathcal{P}$ is only a symmetry if combined either with a discrete
flavor rotation
\begin{equation}
\mathcal{P}^{1,2}_F \colon
\begin{cases}
U_0(x) \rightarrow U_0(x_P), \quad 
U_k(x) \rightarrow U^\dagger_k(x_P - a \hat{k}), 
\quad k = 1, \, 2, \, 3 \\ 
\psi_l(x) \rightarrow i \gamma_0 \tau_{1,2} \psi_l(x_P) \\
\bar{\psi_l}(x) \rightarrow -i \bar{\psi_l}(x_P) \tau_{1,2} \gamma_0  
\end{cases}
\end{equation}
[where $x_P = (-\mathbf{x},t)$, and our generators are normalized
as $\tau_i^2=1$], or combined with
a sign change of the twisted mass term 
\begin{equation}
\widetilde{\mathcal{P}} \equiv \mathcal{P} \times 
[\mu_0 \rightarrow - \mu_0]\,.
\end{equation}
The additional symmetries called $\mathcal{R}_5^{\rm sp}$ and 
$\mathcal{R}_5 \times \mathcal{D}_d$ in Ref.~\cite{FR2003} will
not be needed here.

\subsection{Effective continuum Lagrangian at the quark level}

Following the program of Symanzik~\cite{Symz83a,Symz83b}, the
long distance properties of tmLQCD can be described by an effective continuum
Lagrangian of the form:
\begin{equation} \label{E:SymzL}
\mathcal{L}_{eff} = \mathcal{L}_0 + a \mathcal{L}_1 + a^2
\mathcal{L}_2 + \cdots. 
\end{equation}
The key constraint is that this Lagrangian need only be invariant
under the symmetries of the lattice theory.
We will work through the terms in turn, emphasizing the differences
that are introduced by the twisted mass. The analysis parallels 
and extends that of Ref.~\cite{SS98}.

We consider first $\mathcal{L}_0$, which consists of operators of
dimension four or less, and is the Lagrangian which survives in the
continuum limit. The lattice symmetries restrict its form to be
\begin{align} \label{E:CL0}
\mathcal{L}_{0} =& \; \mathcal{L}_g + \bar{\psi}(D \!\!\!\! / +
Z_m \frac{m_0}{a} + i \gamma_5 \tau_3 \mu) \psi 
- Z_m \frac{\widetilde{m}_c}{a} \bar{\psi}\psi\,,
\end{align}
where $\mathcal{L}_g$ is the continuum gluon Lagrangian,
and, as discussed below, $\mu\propto \mu_0$.
In particular, the parity-flavor symmetry
$\mathcal{P}^{1,2}_F$ requires parity and flavor
to be violated in tandem: it forbids
the flavor singlet parity violating terms $\bar\psi \gamma_5 \psi$
and $\widetilde F_{\mu\nu} F_{\mu\nu}$, as well as the flavor violating,
parity even operator $\bar\psi \tau_3 \psi$.
The residual flavor
$U(1)$ symmetry forbids bilinears containing the flavor matrices $\tau_{1,2}$.
The coefficients $Z_m$, $\mu$ and $m_c$ must be real in order to
retain reflection positivity.
Finally, the spurionic symmetry $\widetilde{\mathcal{P}}$ implies that
the the flavor-parity violating operator $\bar\psi\gamma_5\tau_3\psi$
comes with a coefficient odd in $\mu_0$, here of linear order.

The net effect is that the result (\ref{E:CL0}) differs from that
for Wilson fermions only by the $\mu$ term. In particular, the twisted mass is
only multiplicatively and not additively renormalized~\cite{Frez01}.

The relationship between the parameters in the continuum Lagrangian
and those in the bare lattice Lagrangian has been given in Ref.~\cite{Frez01}.
The continuum fields $\psi$ are related to the
bare lattice fields as follows:
\begin{equation} \label{eq:psirel}
\psi = a^{-3/2}\, Z[g^2(a),\ln(\Lambda_{\rm reg} a)] \, \psi_l   \,,
\end{equation}
where $Z$ is a matching factor, and $\Lambda_{\rm reg}$ is
the renormalization scale of the continuum theory.
The physical quark mass is defined in the usual way,
\begin{equation} \label{eq:mrel}
m = Z_m(m_0 - \widetilde m_c)/a \,,
\end{equation}
while the physical twisted mass is related to the
bare parameters as
\begin{equation} \label{eq:murel}
\mu = Z_\mu \mu_0/a = Z_P^{-1} \mu_0/a \,,
\end{equation}
with $Z_P$ is the matching factor for the pseudoscalar density.
Finally, the coupling constants in the continuum and lattice
theories are related in the standard way.
All the renormalization factors depend on $g^2(a)$ and,
due to their anomalous dimensions, on $\ln(\Lambda_{\rm reg}a)$. We 
keep this (weak) dependence on the lattice spacing implicit.
It should be borne in mind, however, that when,
in the following, we refer to a power-law dependence on $a$,
there will always be implicit sub-leading logarithmic corrections.
We also note that the $Z$ factors are mass independent
(assuming that we are using a mass-independent regularization scheme
in the continuum) and thus are the same for the twisted mass
theory as for normal Wilson fermions. The same is true of $\widetilde m_c$.

In terms of the physical parameters, we finally have
\begin{align} \label{E:CL0final}
\mathcal{L}_{0} =& \; \mathcal{L}_g + \bar{\psi}(D \!\!\!\! / +
m + i \gamma_5 \tau_3 \mu) \psi \,.
\end{align}
In this theory, unlike on the lattice,
the apparent flavor and parity breaking is fake, as $\mu$
can be rotated away by a non-anomalous axial rotation.
This theory is thus equivalent to the usual, untwisted, 
two-flavor QCD with a mass $m' = \sqrt{m^2 + \mu^2}$.

\bigskip

We next construct $\mathcal{L}_1$, which contains dimension
five operators.
The enumeration of these operators is similar to that carried 
out in Ref.~\cite{LSSW96}, and we find
\begin{align} \label{E:CL1}
\mathcal{L}_{1} =& \; 
b_1 \bar{\psi} i \sigma_{\mu\nu} F_{\mu\nu} \psi 
+ b_2 \bar{\psi} (D \!\!\!\! / + m + i \gamma_5 \tau_3 \mu)^2
\psi \notag \\ 
&+ b_3 m \bar{\psi} (D \!\!\!\! / + m + i \gamma_5 \tau_3 \mu)
\psi 
+ b_4 m \mathcal{L}_g 
+ b_5  m^2 \bar{\psi} \psi \notag \\ 
&+ b_6 \mu \bar{\psi} \big\{ (D \!\!\!\! / + m + i \gamma_5
\tau_3 \mu), i \gamma_5 \tau_3 \big\} \psi 
+ b_7  \mu^2 \bar{\psi} \psi 
\end{align}
The coefficients $b_i$ must be real for reflection positivity,
and have a similar implicit weak
dependence on $a$ as do the $Z$ factors discussed above. 
Note that the form of $\mathcal{L}_1$ is similar to that for untwisted Wilson fermions~\cite{SS98},
except for the inclusion of the twisted mass in the
operator $(D \!\!\!\! / + m + i \gamma_5 \tau_3 \mu)$,
and the addition of the $b_6$ and $b_7$ terms.

A number of potential terms have been excluded by the
$\widetilde{\mathcal{P}}$ symmetry: $m\mu\bar\psi\psi$,
$m^2 \bar\psi i\gamma_5\tau_3 \psi$,
$\mu^2 \bar\psi i\gamma_5\tau_3 \psi$,
$\bar\psi D^2 i\gamma_5 \tau_3 \psi$ and
$\bar{\psi} i \sigma_{\mu\nu} F_{\mu\nu} i \gamma_5 \tau_3 \psi$.
The last of these, the ``twisted Pauli term'', requires a factor
of $\mu$ and thus appears only in $\mathcal{L}_2$.
Charge conjugation symmetry enters at $O(a)$, forbidding a
term similar to that with coefficient $b_6$ but containing
a commutator instead of an anticommutator.

We can simplify $\mathcal{L}_{1}$ by dropping terms
that vanish by the leading order equation of motion, i.e.
that which follows from $\mathcal{L}_0$.
This is equivalent to changing quark variables by an amount
proportional to $a$. It removes the $b_2$, $b_3$ and $b_6$ terms.\footnote{%
This is not strictly necessary. Terms vanishing by the equations of
motion break the continuum symmetries in the same way
as other terms which do not vanish. Since it is the symmetry breaking 
properties that matter when constructing the effective 
chiral Lagrangian, the form of the latter is the same whether
or not the vanishing terms are kept. Nevertheless, we drop these
terms as it simplifies the equations.}

The situation with the $b_4$, $b_5$ and $b_7$ terms is more subtle.
On the one hand, they can be removed by $O(a)$ redefinitions
of the parameters in $\mathcal{L}_0$
(the coupling constant for $b_4$, and the untwisted mass for $b_5$ and $b_7$).
Note that for $b_7$ this is an additive redefinition ($m \to m + b_7 a\mu^2$)
which leads to a curvature in the critical mass $m_c$ as a function of
the twisted mass. 
On the other hand, when one makes such a redefinition one loses direct
contact with the underlying bare parameters,
which are linearly related to $g$, $m$ and $\mu$ as explained above.
If one wants to map out
the phase diagram in terms of bare parameters, which is what one does
in a numerical simulation, one should keep the $b_4$, $b_5$ and $b_7$ terms.

At this point, however, it is useful to anticipate the power-counting scheme
that we use in our chiral effective theory. We consider
$m/\Lambda$, $\mu/\Lambda$ and $a \Lambda$
as small quantities of the same size, and work to quadratic order
in a combined chiral-continuum expansion. 
In this power-counting the three terms in question are all of cubic order,
and can be dropped, avoiding the choice discussed in the previous paragraph.
That the $b_5$ and $b_7$ terms are of cubic order is manifest, since they
involve two powers of quark masses in addition to the overall factor of $a$
multiplying $\mathcal{L}_1$. 
The $b_4$ term appears to be only of quadratic order, since it involves
only a single power of the quark mass. However, this is a
 {\em relative} correction to a leading order term,
and the leading order terms in the chiral effective theory are 
themselves of linear order. Thus its contributions to the effective
theory are of cubic order.

\bigskip
Since we are working to quadratic order in the joint chiral-continuum
expansion, we need to determine the form of $\mathcal{L}_2$, which
contains dimension six operators.
There are three such purely gluonic operators~\cite{LW85a,LW85b}.
We do not write these down since,
as explained in Ref.~\cite{LS99}, they will lead to terms of
too high order in our expansion.
In particular, those which do not break the continuum rotation
symmetry give rise to relative corrections proportional to $a^2$.
Since the leading term in the chiral-continuum expansion
is of $O(m,a)$, these corrections are of cubic order and can be dropped.
There are also gluonic operators which break the rotation symmetry down
to its hypercubic subgroup, but these only enter the
chiral Lagrangian at $O(a^2 m^2)$ because of the need to have four derivatives.

The analysis of fermionic operators in
the untwisted Wilson theory has been carried out in Refs.~\cite{SW85,ORS03}.
There are fermion bilinears and four-fermion operators.
What is found, however, is that the only operators which break more
symmetries than those already broken by $\mathcal{L}_0+\mathcal{L}_1$
are the bilinears which break rotation symmetry
down to its hypercubic subgroup. As for the gluonic operators discussed above, 
these contribute beyond quadratic
order in our expansion. The contributions of the remaining operators 
to the effective chiral Lagrangian are of exactly the same form as those
obtained by treating the effects at $\mathcal{L}_1$ at second order.
Thus we do not list these operators.

This leaves the new dimension six operators induced by the twisted mass.
These, however, are all of cubic or higher order in our expansion.
This is because they have at least one factor of $\mu$ in addition to the
$a^2$ common to all operators in $\mathcal{L}_2$. An important example is
the twisted Pauli term discussed above.

\bigskip
Thus we conclude that, for the purpose of constructing the
effective chiral Lagrangian to next-to-leading order (NLO) we
need only keep the following terms at the quark level
\begin{equation} \label{E:CLeff2}
\mathcal{L}_{eff} = \mathcal{L}_g + \bar{\psi}(D \!\!\!\! / 
+ m + i \gamma_5 \tau_3 \mu) \psi 
+ b_1 a \bar{\psi} i \sigma_{\mu\nu} F_{\mu\nu} \psi\,,
\end{equation}
with the relations to bare parameters given in
eqs.~(\ref{eq:psirel}), (\ref{eq:mrel}) and (\ref{eq:murel}).
The only difference from the result for the
standard Wilson theory given in Ref.~\cite{SS98} 
is the addition of a twisted mass term.

\subsection{Effective chiral Lagrangian}

Following \cite{SS98}, the next step is to write down a generalization
of the continuum chiral Lagrangian that includes the effects of the
Pauli term. As already noted, we use the following power 
counting scheme:\footnote{%
Factors of $\Lambda$ needed to make quantities dimensionless
are implicit from now on unless otherwise specified}.
\begin{equation} \label{AokiPC}
1 \gg \{\mathsf{m}, \, p^2,\, a \} \gg
\{\mathsf{m}^2, \, \mathsf{m}p^2, \, p^4, a\mathsf{m}, \,
ap^2, \, a^2 \} \gg \{\mathsf{m}^3, \dots \} \,.
\end{equation}
Here \textsf{m} is a generic mass parameter that can be
either the renormalized normal mass $m$, or the
renormalized twisted mass $\mu$. 
There is no $O(1)$ term in the chiral expansion for pions.
The leading order (LO) terms are of linear order in this expansion,
and the NLO terms of quadratic order. We work to NLO, which is sufficient
for the study of the phase diagram.
We stress that we can use the expansion in other regimes,
e.g. $\mathsf{m} \gg a$ or $\mathsf{m} \sim a^2$, except that we
then need to drop certain terms which become of too high order.

The chiral Lagrangian is built from the $SU(2)$ matrix-valued $\Sigma$
field, which contains the relevant low-energy degrees of
freedom. It transforms under the chiral group $SU(2)_L \times SU(2)_R$
as 
\begin{equation}
\Sigma \rightarrow L \Sigma R^{\dagger}, \qquad L \in SU(2)_L, \; R
\in SU(2)_R. 
\end{equation}
The vacuum expectation value of $\Sigma_0$ breaks the chiral symmetry
down to an $SU(2)$ subgroup. The fluctuations around $\Sigma_0$
correspond to the pseudo-scalar mesons 
(pions): 
\begin{equation} \label{E:Sfield}
\Sigma(x) = \Sigma_0 \exp \bigg\{i \sum_{a = 1}^3 \pi_a(x) \tau_a / f
\bigg\} = \Sigma_0 \big[ \cos(\pi(x)/f) + i \hat{\boldsymbol{\pi}}(x)
\cdot \boldsymbol{\tau} \sin(\pi(x)/f) \big], 
\end{equation}
where $f$ is the decay constant (normalized so that $f_\pi=93\;$MeV).
The norm and the unit vector of the pion fields are given by
$\pi = \sqrt{\boldsymbol{\pi} \cdot \boldsymbol{\pi}} = \pi_a \pi^a$
and $\hat{\pi}_a = \pi_a / \pi$. 

The chiral Lagrangian can be obtained from the quark Lagrangian,
eq.~(\ref{E:CLeff2}), by a standard spurion analysis. 
We must introduce a spurion matrix $\hat A$ for the Pauli term,
as well as the usual spurion $\chi$ for the mass terms.
Both transform in the same way as the $\Sigma$ field.
At the end of the analysis the spurions are set to their
respective constant values
\begin{equation}
\chi \longrightarrow 2 B_0 (m + i \tau^3 \mu)
                   \equiv \hat{m} + i \tau^3 \hat{\mu} 
\qquad \mathrm{and} \qquad 
\hat A \longrightarrow 2 W_0 \, a \equiv \hat{a} \,,
\end{equation}
where $B_0$ and $W_0$ are unknown dimensionful parameters,
and we have defined useful quantities $\hat m$, $\hat \mu$ and $\hat a$.
Note that the only change caused by the presence of the twisted mass is
the appearance of the $\mu$ term in the constant value of $\chi$.
Other than this, the construction of the chiral Lagrangian is
identical to that for untwisted Wilson fermions.

Because of this simplification, we can read off the form
of the chiral Lagrangian for tmLQCD from Ref.~\cite{ORS03},
in which  the Lagrangian for untwisted Wilson fermions
was worked out to quadratic order in our expansion.
The only extension we make is to include sources for currents
and densities. The left- and right-handed
currents  are introduced in the standard way by using
the covariant derivative $D_\mu \Sigma =
\partial_\mu \Sigma - i l_\mu \Sigma + i \Sigma r_\mu$, 
and the associated field strengths, 
e.g. $L_{\mu\nu}=\partial_\mu l_\nu - \partial_\nu l_\mu + i [l_\mu,l_\nu]$,
and enforcing invariance under 
local chiral transformations~\cite{GassLeut84,GassLeut85}.
Sources for scalar and pseudoscalar densities are similarly included
if we write $\chi = 2B_0 (s + ip)$, with 
$s$ and $p$ hermitian matrix fields.
Although the sources play no role in the discussion of the
phase diagram, we have used them 
to obtain forms for various matrix elements of phenomenological
interest~\cite{Wulat04}. These results will be described elsewhere~\cite{SWPrep}.

Putting these ingredients together the resulting effective chiral
Lagrangian is (in  Euclidean space)
\begin{align} \label{E:ChL}
\mathcal{L}_\chi &= 
 \frac{f^2}{4} \mathrm{Tr}(D_\mu \Sigma D_\mu \Sigma^\dagger)
-\frac{f^2}{4} \mathrm{Tr}(\chi^{\dagger} \Sigma + \Sigma^\dagger\chi) 
-\frac{f^2}{4} \mathrm{Tr}(\hat{A}^{\dagger} \Sigma + 
               \Sigma^\dagger\hat{A}) \notag \\ 
&\quad
- L_1 \mathrm{Tr}(D_\mu \Sigma D_\mu \Sigma^\dagger)^2
- L_2 \mathrm{Tr}(D_\mu \Sigma D_\nu \Sigma^\dagger)
      \mathrm{Tr}(D_\mu \Sigma D_\nu \Sigma^\dagger) \notag \\
&\quad 
+ (L_4 + L_5/2)\mathrm{Tr}(D_\mu \Sigma^\dagger D_\mu \Sigma)
      \mathrm{Tr}(\chi^{\dagger} \Sigma +  \Sigma^\dagger\chi)
\notag \\ &\quad
+ L_5 \left\{
\mathrm{Tr}\left[(D_\mu \Sigma^\dagger D_\mu \Sigma)
      (\chi^{\dagger} \Sigma +  \Sigma^\dagger\chi)\right]
-\mathrm{Tr}(D_\mu \Sigma^\dagger D_\mu \Sigma)
      \mathrm{Tr}(\chi^{\dagger} \Sigma +  \Sigma^\dagger\chi)/2 \right\}
\notag \\ &\quad
- (L_6 + L_8/2)  
\big[\mathrm{Tr}(\chi^{\dagger} \Sigma + \Sigma^\dagger\chi)\big]^2
- L_8 \left\{
\mathrm{Tr}\big[(\chi^{\dagger} \Sigma + \Sigma^\dagger\chi)^2\big]
-
\big[\mathrm{Tr}(\chi^{\dagger} \Sigma + \Sigma^\dagger\chi)\big]^2/2
\right\}
\notag \\ &\quad
- L_7 \big[\mathrm{Tr}(\chi^{\dagger} \Sigma - \Sigma^\dagger\chi)\big]^2
\notag \\ &\quad
+ i L_{12} \mathrm{Tr}(L_{\mu\nu} D_\mu \Sigma D_\nu \Sigma^\dagger +
           R_{\mu\nu} D_\mu \Sigma^\dagger D_\nu \Sigma)
+ L_{13} \mathrm{Tr}(L_{\mu\nu} \Sigma R_{\mu\nu} \Sigma) \notag \\
&\quad
+ (W_4 + W_5/2)\mathrm{Tr}(D_\mu \Sigma^\dagger D_\mu \Sigma)
      \mathrm{Tr}(\hat{A}^{\dagger} \Sigma + \Sigma^{\dagger}\hat{A})
- (W_6 + W_8/2)\mathrm{Tr}(\chi^{\dagger} \Sigma + \Sigma^\dagger\chi) 
      \mathrm{Tr}(\hat{A}^{\dagger} \Sigma + \Sigma^{\dagger}\hat{A})
      \notag \\
&\quad
+ W_{10} \mathrm{Tr}(D_\mu \hat{A}^\dagger D_\mu \Sigma + 
          D_\mu \Sigma^\dagger D_\mu \hat{A})
- (W_6'+W_8'/2) \big[\mathrm{Tr}(\hat{A}^{\dagger} \Sigma +
       \Sigma^{\dagger}\hat{A})\big]^2 \notag \\
&\quad
+ \mathrm{contact \; terms} \,.
\end{align}
Here the $L_i$'s
are the standard Gasser-Leutwyler low-energy constants of continuum 
chiral perturbation theory,
and the $W_i$'s and $W'_i$'s unknown low-energy
constants associated with discretization errors.
For the latter constants we use the notation of Ref.~\cite{RS02,ORS03}.
In fact, these constants related to discretization
are the same as those for the untwisted
Wilson theory, assuming the same gauge action, 
since the twisting enters only through the parameters in the spurions.

In writing (\ref{E:ChL}) 
we have used various simplifications
that result from the fact that the flavor group is $SU(2)$.
In particular, because
 $\hat A$ is, in the end, proportional to an element of $SU(2)$
with a real proportionality constant, and because we do not
use it as a source, we can take 
$\hat{A}^{\dagger}\Sigma+\Sigma^{\dagger}\hat{A}$ to
be proportional to the identity matrix,
and $\mathrm{Tr}(\hat{A}^{\dagger}\Sigma-\Sigma^{\dagger}\hat{A})$
to vanish.
Thus the following possible terms vanish for $SU(2)$:
\begin{align}
& +W_5 \left\{
\mathrm{Tr}\big[(D_\mu \Sigma^\dagger D_\mu \Sigma)
      (\hat{A}^{\dagger} \Sigma + \Sigma^{\dagger}\hat{A})\big]
-\mathrm{Tr}(D_\mu \Sigma^\dagger D_\mu \Sigma)
      \mathrm{Tr}(\hat{A}^{\dagger} \Sigma + \Sigma^{\dagger}\hat{A})/2
\right\}
\notag \\ &
-W_8 \left\{
\mathrm{Tr}\big[(\chi^{\dagger} \Sigma +  \Sigma^\dagger\chi) 
      (\hat{A}^{\dagger} \Sigma + \Sigma^{\dagger}\hat{A})\big]
- \mathrm{Tr}(\chi^{\dagger} \Sigma + \Sigma^\dagger\chi) 
      \mathrm{Tr}(\hat{A}^{\dagger} \Sigma + \Sigma^{\dagger}\hat{A})/2
\right\}
\notag \\ &
-W_7 
\mathrm{Tr}(\chi^{\dagger} \Sigma - \Sigma^\dagger\chi) 
\mathrm{Tr}(\hat{A}^{\dagger} \Sigma - \Sigma^{\dagger}\hat{A})
\notag \\ &
-W_8' \left\{
\mathrm{Tr}\big[(\hat{A}^{\dagger} \Sigma + \Sigma^{\dagger}\hat{A})^2 \big]
-
\big[\mathrm{Tr}(\hat{A}^{\dagger} \Sigma +\Sigma^{\dagger}\hat{A})\big]^2/2
\right\} 
\notag \\ &
-W_7'
\big[\mathrm{Tr}(\hat{A}^{\dagger} \Sigma - \Sigma^{\dagger}\hat{A})\big]^2
\,.
\end{align}
Similarly, the operators multiplied by $L_5$, $L_7$ and $L_8$ alone in (\ref{E:ChL})
do not contribute once we set $\chi$ to 
$2 B_0 (m + i \tau^3 \mu)$, although they are needed when using
$\chi$ as a source. They do not contribute in our subsequent study
of the phase diagram.

Finally, we comment on the $W_{10}$ term, which is present
only because we include both discretization errors and external sources.
This term can, in fact, be removed using the equations of motion,
but we have found that it provides a useful diagnostic in computations of
matrix elements.

Our result (\ref{E:ChL}) agrees with, and extends, the work of Ref.~\cite{MS04},
in which the chiral Lagrangian for tmLQCD including effects linear in
$a$ was determined. Our generalizations are to include the term of $O(a^2)$,
to make the simplifications due to using the group $SU(2)$, and to
include the sources for currents.
In comparing our result to that in Ref.~\cite{MS04} it should be noted
that the result in that work is expressed in the physical basis,
while we use the twisted basis. The two results are related by a non-anomalous
axial rotation.

\section{\label{sec:anaAoki} Analysis of phase diagram}

To investigate the phase diagram of tmLQCD, we are
interested in the vacuum state of the effective continuum chiral
theory. To the order that we are working, the potential energy is
\begin{align} \label{E:PE}
V_{\chi} = &
-\frac{c_1}{4} \mathrm{Tr}(\Sigma + \Sigma^{\dagger}) 
+ \frac{c_2}{16} \left[\mathrm{Tr}(\Sigma + \Sigma^{\dagger})\right]^2
+ \frac{c_3}{4} \mathrm{Tr} \big[i(\Sigma - \Sigma^{\dagger}) \tau_3
\big] \notag \\
&+ \frac{c_4}{16} \left\{\mathrm{Tr}\big[i(\Sigma - \Sigma^{\dagger}) \tau_3
\big]\right\}^2
+ \frac{c_5}{16} \mathrm{Tr} \big[i(\Sigma - \Sigma^{\dagger}) 
\tau_3\big] \mathrm{Tr}(\Sigma + \Sigma^{\dagger}), 
\end{align}
where the explicit forms of the coefficients,
 and their sizes in our power-counting scheme, are
\begin{align} \label{E:ChLcoefs}
c_1 &= f^2(\hat{m} + \hat{a}) \sim m + a,
\notag \\ 
c_2 &= -8 \left[(2L_6 + L_8)\hat{m}^2
 + (2W_6+ W_8)\, \hat{a} \, \hat{m} + (2W_6'+W_8') \, \hat{a}^2 \right] 
\sim m^2 + a m  + a^2 , \notag \\
c_3 &= f^2 \hat{\mu} \sim \mu, \notag \\
c_4 &= -8 (2L_6 + L_8)\, \hat{\mu}^2 \sim \mu^2, \notag \\
c_5 &= 16 \left[(2L_6+L_8) \, \hat{m} \hat{\mu} +
(W_6+ W_8/2) \, \hat{a} \, \hat{\mu} \right] \sim \mu ( m + a) \,.
\end{align}
Note that there are no relations between the $c_i$ coefficients: all
five are independent at non-zero lattice spacing.
The extra terms introduced by twisting are those with coefficients
$c_3$, $c_4$ and $c_5$.

To determine the pattern of symmetry breaking as a function of the
coefficients $c_i$, we repeat the analysis of
\cite{SS98}. We distinguish three regions of quark masses, each
successively smaller by a factor of $a$: 

\smallskip\noindent
(i) {\bf Physical quark masses:} $1 \gg m \sim \mu \gg a$. 
In this case, both discretization errors and
terms of quadratic order in masses can be neglected, so that only the
$c_1\sim m$ and $c_3\sim \mu$ terms need be kept.
In this case the symmetry breaking is as in the continuum.
In particular, the condensate $\Sigma_0$ lies in the direction of
the full mass term, so that:
\begin{equation}
\Sigma_0 = \frac{(m + i \tau_3 \mu)}{\sqrt{m^2+\mu^2}}
= \frac{(\hat m + i \tau_3 \hat\mu)}{\sqrt{\hat m^2+\hat \mu^2}}
 \end{equation}
The pions are degenerate, with masses
\begin{equation}
m_\pi^2 = \sqrt{\hat m^2 + \hat \mu^2} \left[1 + O(a,m,\mu)\right]
\,.
\end{equation}
Note that this result holds also if either $m\gg a$ or $\mu\gg a$.

\smallskip\noindent
(ii) {\bf Significant discretization errors:} $1 \gg  m \sim \mu \sim a$.
This is the parameter region for which our expansion is most natural.
The LO coefficients $c_{1,3}$ still dominate over the NLO coefficients
$c_{2,4,5}$, but the $O(a)$ term in $c_1$ cannot be ignored.
Thus the results for region (i) still hold, except that the untwisted quark
mass $\hat m$ must be replaced by the shifted mass $\hat m'=\hat m + \hat a$,
or equivalently $m \to m'=m + a W_0/B_0$.
In other words, there is an $O(a)$ shift in the critical mass,
as noted in Ref.~\cite{SS98}.
In terms of this new mass the coefficients become  
\begin{align}
c_1 &\sim m', \quad
c_2 \sim m'^2 + a m' + a^2 , \quad
c_3 \sim \mu, \quad
c_4 \sim \mu^2, \quad
c_5 \sim \mu(m' + a).
\end{align}

\smallskip\noindent
(iii) {\bf Aoki region:} $m' \sim \mu\sim a^2$. 
In this region the nominally NLO coefficient $c_2$ is of
the same size as the LO coefficients $c_{1,3}$; all are of $O(a^2)$.
The other NLO coefficients remain suppressed by at least one
power of $a$:  $c_{4}\sim a^4$, $c_5\sim a^3$.
The fact that LO and some NLO coefficients are comparable might
suggest a breakdown in convergence. This is not so, however.
The largest NNLO term is this parameter region is $\sim a^3$,
and thus suppressed by one power of $a$.
Loop corrections are also suppressed, since they are
quadratic in $m'$ and $\mu$ (up to logarithms) and thus $\sim a^4$.
The key result that allows this reordering of the series is
that the leading order discretization error has exactly the same
form as the term proportional to $m$ and so can be completely absorbed
into $m'$, to all orders in the chiral expansion.

As in the untwisted Wilson theory, it is the competition between
the LO terms and the NLO $c_2$ term that can lead to interesting
phase structure.
Note that {\em both} $m'$ and $\mu$ must be of $O(a^2)$ in order 
for such competition to occur;
if either $m'$ or $\mu$ is of $O(a)$ then one
is in the continuum-like region (ii).

\bigskip
To determine the condensate we must minimize the potential energy (\ref{E:PE}). 
We parameterize the chiral field in the standard way:
 $\Sigma = A + i\mathbf{B} \cdot \boldsymbol{\tau}$ with 
real $A$ and $\mathbf{B}$ satisfying $A^2 + \mathbf{B}^2 = 1$,
so that  $A, \, B_i \in [-1,1]$.
Similarly, the condensate (the value of $\Sigma$ that
minimizes the potential) is written 
$\Sigma_0 = A_m + i\mathbf{B}_m \cdot \boldsymbol{\tau}$.
In the Aoki region the potential is\footnote{%
Henceforth, we will drop references to the $O(a^3)$ corrections.}
\begin{equation} \label{E:PE2}
V_{\chi} = -c_1 A - c_3 B_3 + c_2 A^2\,.
\end{equation}
Although $c_1=f^2 \hat m'$ and $c_3=f^2 \hat \mu$ can both
take either sign, we need only consider
the case when both are positive. The other possibilities
can be obtained using the symmetries of $V_\chi$:
if $c_1\to -c_1$ with $c_{2,3}$ fixed
(i.e. when $m'\to -m'$), the condensate changes 
as $A_m \to - A_m$, $B_{3,m} \to B_{3,m}$;
while if $c_3 \to - c_3$ with $c_{1,2}$ fixed
(i.e. when $\mu\to-\mu$), then
$A_m \to A_m$, $B_{3,m} \to - B_{3,m}$.
We do not need to specify the transformation
of the other two components of the condensate
since they vanish, as we now show.

To do so it is useful to define $r$ and $\theta$ by
\begin{equation}
A = r \cos\theta,\ \ B_3 = r \sin\theta,\ \
A^2+B_3^2=r^2, \ \ B_1^2+B_2^2= 1-r^2,\ \ 0\le r \le 1\,,
\end{equation}
in terms of which the potential becomes
\begin{equation}
V_\chi = -r (c_1 \cos\theta +  c_3 \sin\theta) + r^2 c_2 \cos^2\theta
\,.
\end{equation}
We will first minimize this at fixed $r$, and then minimize
with respect to $r$.
Based on the symmetries just discussed, we consider only
$c_{1,3}>0$. At fixed $r$,
the term linear in $r$, $-(c_1 \cos\theta +  c_3 \sin\theta)$,
has its minimum in the quadrant $0<\theta<\pi/2$,
while the term quadratic in $r$, $c_2\cos^2\theta$, has
its minima at $\pm \pi/2$ for $c_2>0$ and
$0,\pi$ for $c_2<0$. It follows that, irrespective
of the sign of $c_2$, the minimizing angle, $\theta_0(r)$,
also lies in the first quadrant: $0<\theta_0(r)<\pi/2$.
Its actual value is given by the appropriate solution to
\begin{equation}
\label{eq:theta0r}
c_1 \sin \theta_0(r) - c_3 \cos\theta_0(r)
= c_2 r \sin[2\theta_0(r)] \,,
\end{equation}
and we denote the value of the potential at
the minimum by $V_{\chi,min}(r)$.
To minimize with respect to $r$ we evaluate the derivative:
\begin{equation}\label{eq:dVdr}
\frac{d V_{\chi,min}(r)}{dr} = 
\left.\frac{\partial V_{\chi}}{\partial r}\right|_{\theta=\theta_0(r)}
= - \frac{c_3}{\sin\theta_0(r)} \,,
\end{equation}
where to obtain this simple form we have used eq.~(\ref{eq:theta0r}).
Since $c_3>0$ by assumption, and $\sin\theta_0(r)>0$
as $\theta_0(r)$ lies in the first quadrant, we find that
$V_{\chi,min}(r)$ is a monotonically decreasing function of $r$.
The symmetries discussed above show that this holds for
any $c_{1,3}$.
Thus the absolute minimum of $V_\chi$ is at $r=1$,
and the minimizing angle is $\theta_m\equiv\theta_0(r=1)$.

At this point it is useful to introduce scaled variables
\begin{align}
\alpha &= \frac{c_1}{|c_2|} = 
\frac{f^2 (\hat m + \hat a)}{8|2W_6'+W_8'| \hat a^2}
\sim m'/a^2 
\label{eq:alphadef} \\
\beta &= \frac{c_3}{|c_2|} = 
\frac{f^2 \hat \mu}{8|2W_6'+W_8'| \hat a^2} \sim \mu/a^2 \,.
\label{eq:betadef}
\end{align}
In words, $\alpha$ is the shifted, untwisted mass in units proportional
to $a^2$, while $\beta$ is the twisted mass in the same units.
In the Aoki region $\alpha$ and $\beta$ are of order unity.
In terms of these variables, the equation to be solved to
determine $A_m=\cos\theta_m$  is
\begin{equation} \label{E:VminPoly}
A_m^4 \mp \alpha A_m^3 +\frac{\alpha^2 + \beta^2 - 4}{4}A_m^2 
\pm \alpha A_m - \frac{\alpha^2}{4} = 0\,.
\end{equation} 
[This is just eq.~(\ref{eq:theta0r}) for $r=1$ after some manipulation.]
Here the upper sign is for $c_2>0$, the lower for $c_2<0$.
One must pick solutions with $A_m$ real 
and satisfying $-1\le A_m \le 1$, and of these choose that
which minimizes $V_\chi$. Given $A_m$, the other components
are given by
\begin{equation}
B_{1,m}=B_{2,m}=0\,,\qquad B_{3,m} =  \mathrm{sign}(c_3) \sqrt{1 - A_m^2}
\,,
\end{equation}
where the sign of $B_{3,m}$ follows from the form of the
potential, eq.~(\ref{E:PE2}).

The pion masses are given by the quadratic fluctuations about
the condensate, using eq.~(\ref{E:Sfield}).  
The unbroken $U(1)$ flavor symmetry ensures that the charged pions
are degenerate. We find
\begin{align} \label{eq:mpi12}
m_{\pi_1}^2 &= m_{\pi_2}^2  
= \frac{c_1 A_m + c_3 B_{3,m} - 2 c_2 A_m^2}{f^2}
= \frac{|c_2|}{f^2} \frac{\beta}{B_{3,m}} 
\\
\label{eq:mpisplit}
\Delta m_\pi^2 &\equiv m_{\pi_3}^2 - m_{\pi_1}^2 = \frac{2 c_2 B_{3,m}^2}{f^2} \,,
\end{align}
where to obtain the final form for charged pions we
have used eq.~(\ref{eq:theta0r}). 
It is straightforward to show that the none of the pion masses
become negative for any values of the parameters.

Before we show plots of the results, it is useful to
discuss how the sign of $c_2$ affects the solutions.
Recall that the sign of $c_2$ 
depends on the gauge action, and is not known \textit{\`{a} priori}.
For the untwisted Wilson action, the sign has an important
impact: the Aoki phase appears only for
$c_2>0$, while there is a first order transition for $c_2<0$.
Once we extend the theory into the full twisted mass plane, however,
the two possibilities are related.
We focus on the $A-B_3$ plane (i.e. we set $r=1$) since we know from above
that the minimum of the potential lies in this plane for all $c_2$.
The potential in this plane is
\begin{equation}
V_\chi = - c_1 \cos\theta - c_3 \sin\theta + c_2 \cos^2\theta \,.
\end{equation}
If we change variables as follows
\begin{equation}
c_1 = c'_3\,,\quad
c_3 = - c'_1\,,\quad
c_2 = - c'_2\,,\quad
\theta = \theta'-\pi/2\,,
\end{equation}
so that $\cos\theta=\sin\theta'$ and $\sin\theta=-\cos\theta'$, then
the potential has the same form
\begin{equation}
V_\chi = -c'_1 \cos\theta' - c'_3 \sin\theta' + c'_2 \cos^2\theta' - c'_2\,,
\end{equation}
aside from an overall shift.
This implies that if we take the phase diagram for, say $c_2>0$, and rotate
it anticlockwise by $90^\circ$, we will obtain the phase diagram for
$c'_2 = - c_2 <0$, with the components of the condensate given by
$A'_m = -B_{3,m}$ and $B'_{3,m} = A_m$. The mass of $\pi_3$ is given 
simply by its value at the rotated point, $m'_{\pi_3} = m_{\pi_3}$.

This discussion does not include fluctuations in the other two directions.
To obtain the charged pion masses, however, we can use the general result
eq.~(\ref{eq:mpisplit}) for the splitting between charged and neutral pion masses,
and the equality of the neutral masses, to obtain
\begin{align}
\left(m'_{\pi_1}\right)^2 &= m_{\pi_1}^2 + \Delta m_\pi^2 - \Delta {m'_\pi}^2
\\
&= m_{\pi_1}^2 + \frac{2 c_2 \sin^2\theta_m}{f^2} -
                 \frac{2 c'_2 \sin^2\theta'_m}{f^2}
\\
&= m_{\pi_1}^2 + \frac{-2 c'_2 \cos^2\theta'_m}{f^2} -
                 \frac{2 c'_2 \sin^2\theta'_m}{f^2}
\\
&= m_{\pi_1}^2 - \frac{2 c'_2}{f^2}\,.
\end{align}
Thus we find a fixed offset (independent of $c_1$ and $c_3$) between
the charged pion masses. The offset has the correct sign so that all
pions have positive or zero mass-squared. In particular, if $c_2>0$
there is an Aoki phase with $m_{\pi_1}=0$. This rotates
(as shown in Fig.~\ref{fig:mmu:b}) into a phase with  
${m'_{\pi_1}}^2= -2 c'_2/f^2 = + 2 c_2/f^2 >0$.

\bigskip

It is straightforward to solve the quartic equation and 
determine the condensate and pion masses. 
The only subtlety is that care must be taken when either $\beta$
or $\alpha$ is set to zero. This can be seen, for example,
from the final form for the charged pion masses, eq.~(\ref{eq:mpi12}),
which becomes $0/0$ when $\beta\to0$. In either of these two limits,
$\alpha\to0$ or $\beta\to0$, the quartic equation reduces to a quadratic.
The solutions for $\beta\to0$ have been given in Ref.~\cite{SS98},
and we do not repeat them here.
Those for $\alpha\to0$ can be obtained from those for $\beta\to0$
using the $90^\circ$ rotation in the mass plane just discussed. 

For the remainder of this article we illustrate 
the nature of the phase diagram by plotting the condensate and pion masses.
We display results as a function of $\alpha$ for fixed $\beta$ and $c_2$.
We only show results for $\beta \ge 0$ since,
as explained above, those for $\beta < 0$
differ only by changing the sign of $B_{3,m}$. In particular, the pion masses
are symmetrical under $\beta \to -\beta$. We could also consider only
$\alpha\ge 0$, but find it clearer to show results for the both signs.

\begin{figure}
\centering
\psfrag{A0}[][]{\large $A_m$}
\psfrag{a}{\large $\alpha$}
\psfrag{b0}{\tiny $\beta = 0$}
\psfrag{b1}{\tiny $\beta = 1$}
\psfrag{b2}{\tiny $\beta = 2$}
\psfrag{b3}{\tiny $\beta = 3$}
\includegraphics[width=3in]{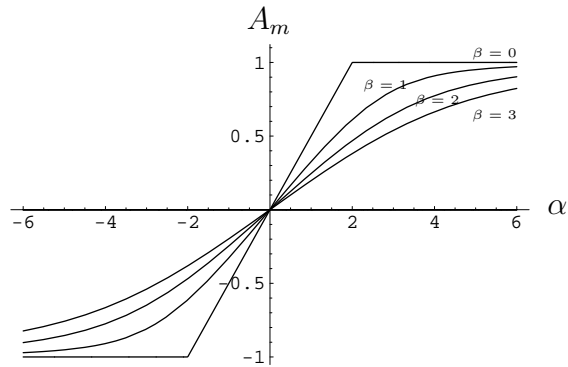}
\caption{\label{fig:c2gt0Am} The global minimum of the potential,
  $A_m$, as a function of $\alpha$, for $c_2 > 0$ and $\beta =
  0,\,1,\,2,\,3$.}
\end{figure}

\begin{figure}
\centering
\subfigure[Mass of $\pi_1$ and $\pi_2$]{
\psfrag{m2f2c2}[][]{\large $m^2_\pi f^2 / c_2$}
\psfrag{a}{\large $\alpha$}
\psfrag{b0}{\small $\beta = 0$}
\psfrag{b1}{\small $\beta = 1$}
\psfrag{b2}{\small $\beta = 2$}
\psfrag{b3}{\small $\beta = 3$}
\label{fig:c2gt0mpi:a}
\includegraphics[width=3in]{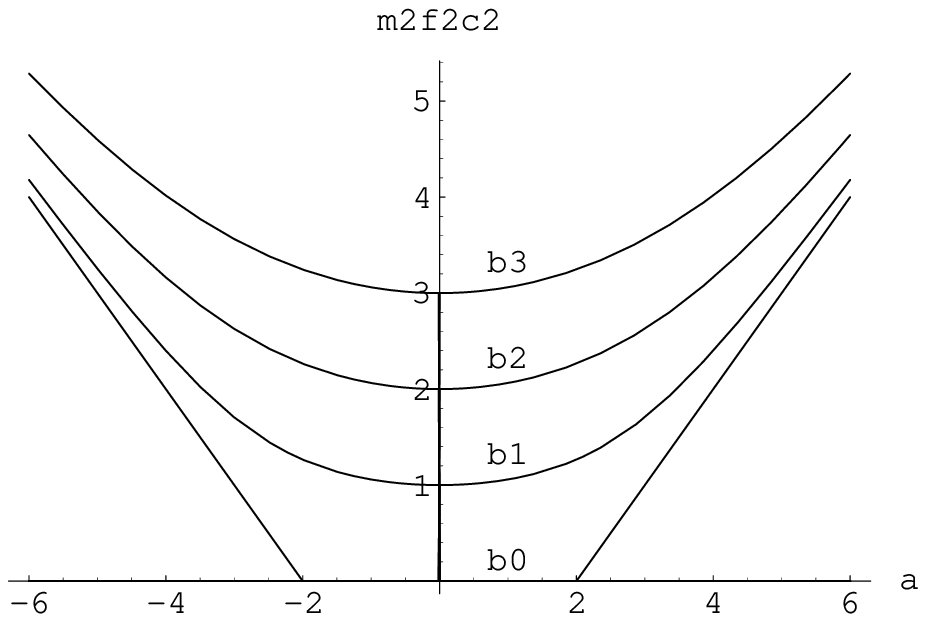}}
\hspace{0.1in}
\subfigure[Mass of $\pi_3$]{
\psfrag{m2f2c2}[][]{\large $m^2_\pi f^2 / c_2$}
\psfrag{a}{\large $\alpha$}
\psfrag{b0}{\small $\beta = 0$}
\psfrag{b1}{\small $\beta = 1$}
\psfrag{b2}{\small $\beta = 2$}
\psfrag{b3}{\small $\beta = 3$}
\label{fig:c2gt0mpi:b}
\includegraphics[width=3in]{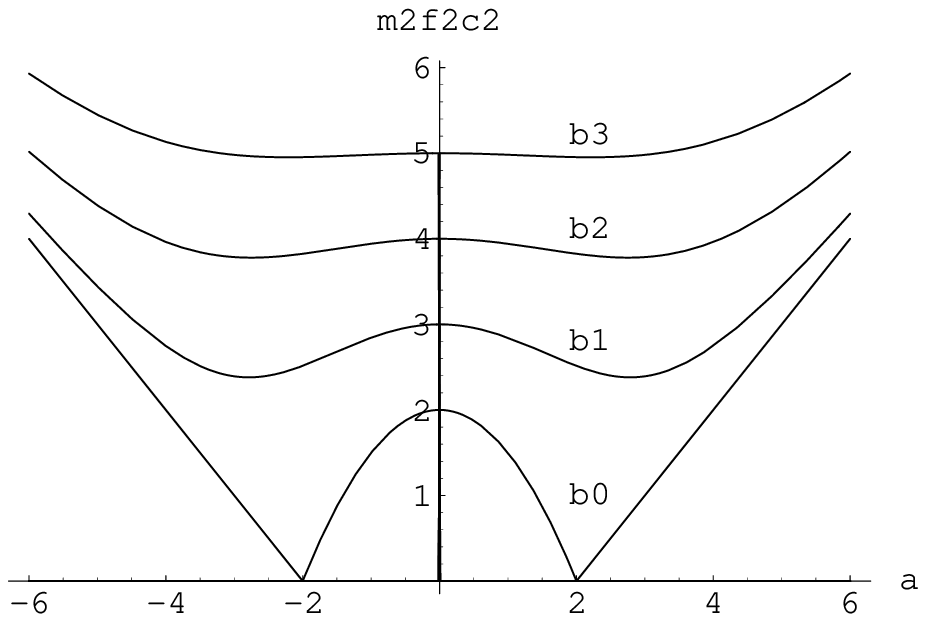}}
\caption{\label{fig:c2gt0mpi} Mass of the pions as a function of
  $\alpha$, for $c_2 > 0$ and $\beta = 0,\,1,\,2,\,3$.} 
\end{figure}

We begin with results for $c_2 > 0$. 
Figure~\ref{fig:c2gt0Am} shows the form of the 
identity component of the condensate, $A_m=\mathrm{Tr}(\Sigma_0)/2$,
for $\beta=0$, $1$, $2$ and $3$.
The corresponding pion masses are shown in Fig.~\ref{fig:c2gt0mpi}.
In the untwisted theory ($\beta=0$)
there are second order transitions at $\alpha=\pm 2$,
as shown by the kinks in $A_m$ and the vanishing of the
pion masses. The Aoki phase, with $B_{3,m}\ne 0$, which breaks flavor
and parity, and correspondingly has two Goldstone bosons, 
lies between these second-order points.
Once $\beta$ is non-vanishing, however, the transition is smoothed
out into a crossover, and the pion masses are always non-zero.
Flavor is broken for all $\alpha$, with the charged pions heavier
than the neutral pion by $O(a^2)$, as given by eq.~(\ref{eq:mpisplit}).
For the special case of $\alpha=0$ (maximal twisting), $A_m$ vanishes
and $B_{3,m}^2=1$, so the mass-squared splitting is 
$2 c_2/f^2$ for all $\beta$ (which becomes a difference of
2 in the units in the plots).

It is now possible to pass through the Aoki phase by varying $\beta$,
i.e. by changing the sign of the twisted mass. When doing so there is
a first-order phase transition, since $B_{3,m}$ 
jumps from $+\sqrt{1-A_m^2}=\sqrt{1-\alpha^2/4}$ 
to $-\sqrt{1-A_m^2}=-\sqrt{1-\alpha^2/4}$.

\bigskip

\begin{figure}
\centering
\subfigure[Global minimum, $\beta = 0$]{
        \psfrag{A0}[][]{\small $A_m$}
        \psfrag{a}{\small $\alpha$}
        \label{fig:c2lt0:a}
        \includegraphics[width=2.2in]{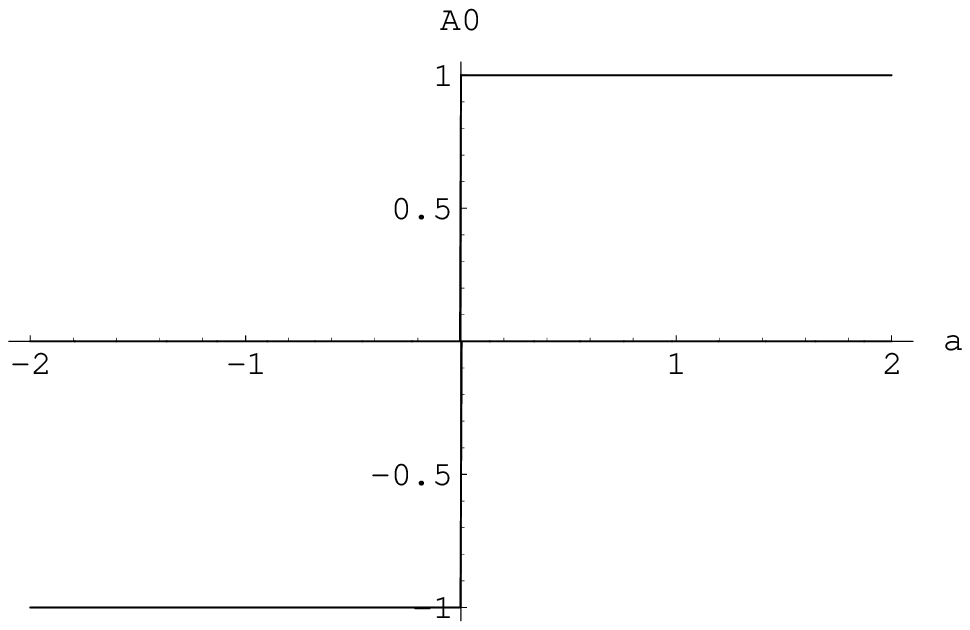}}
\hspace{0.1in}
\subfigure[Pion masses, $\beta = 0$]{
        \psfrag{m2f2c2}[][]{\small $m^2_\pi f^2 / |c_2|$}
        \psfrag{a}{\small $\alpha$}
        \label{fig:c2lt0:b}
        \includegraphics[width=2.2in]{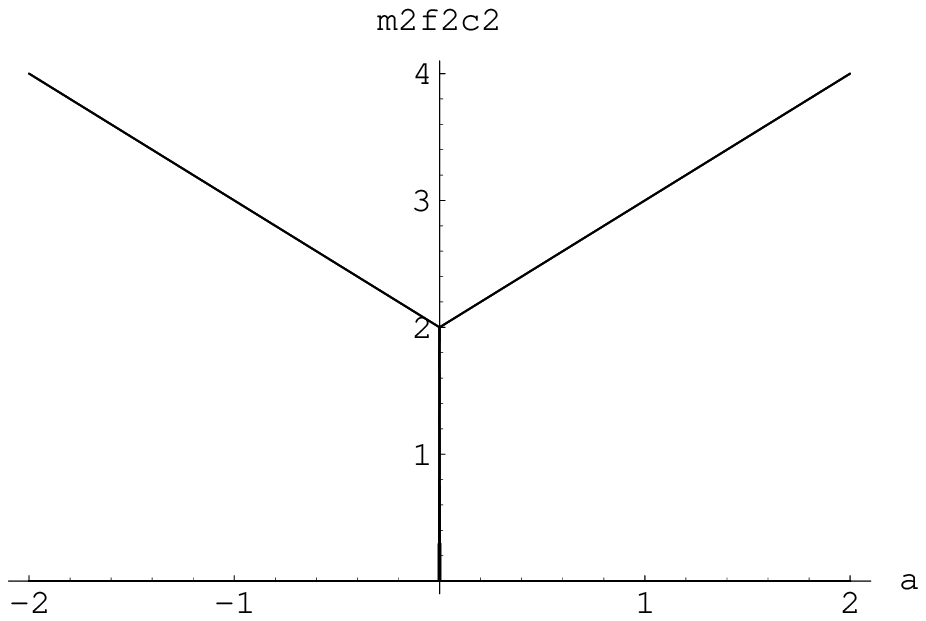}}

\subfigure[Global minimum, $\beta = 1$]{
        \psfrag{A0}[][]{\small $A_m$}
        \psfrag{a}{\small $\alpha$}
        \label{fig:c2lt0:c}
        \includegraphics[width=2.2in]{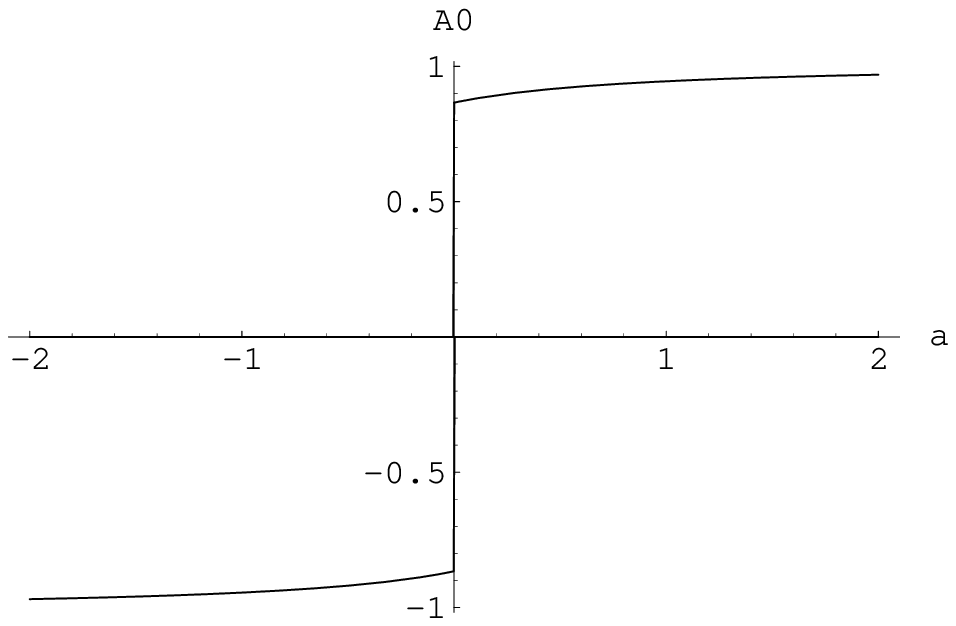}}
\hspace{0.1in}
\subfigure[Pion masses, $\beta = 1$]{
        \psfrag{m2f2c2}[][]{\small $m^2_\pi f^2 / |c_2|$}
        \psfrag{a}{\small $\alpha$}
        \label{fig:c2lt0:d}
        \includegraphics[width=2.2in]{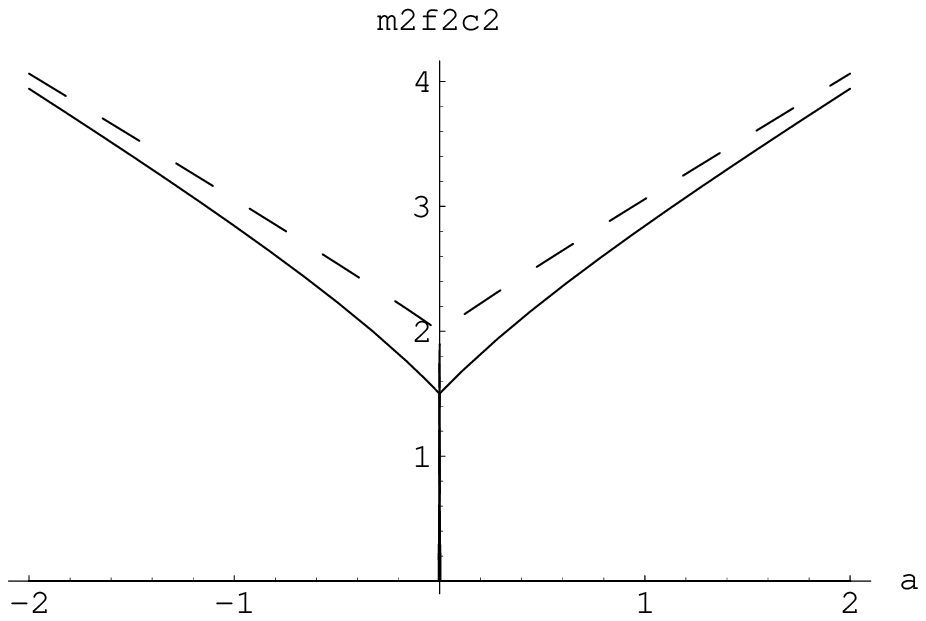}}

\subfigure[Global minimum, $\beta = 2$]{
        \psfrag{A0}[][]{\small $A_m$}
        \psfrag{a}{\small $\alpha$}
        \label{fig:c2lt0:e}
        \includegraphics[width=2.2in]{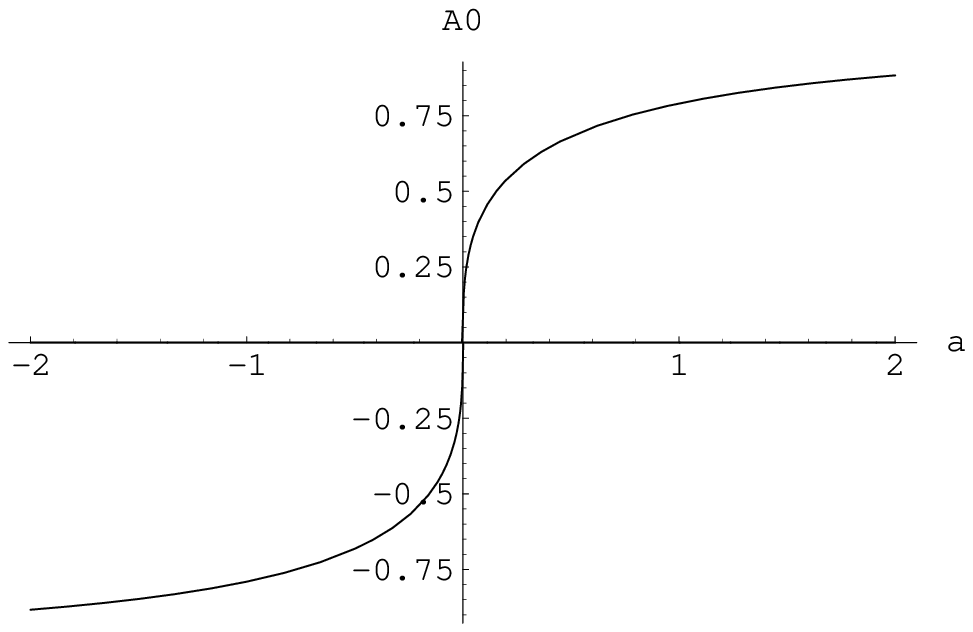}}
\hspace{0.1in}
\subfigure[Pion masses, $\beta = 2$]{
        \psfrag{m2f2c2}[][]{\small $m^2_\pi f^2 / |c_2|$}
        \psfrag{a}{\small $\alpha$}
        \label{fig:c2lt0:f}
        \includegraphics[width=2.2in]{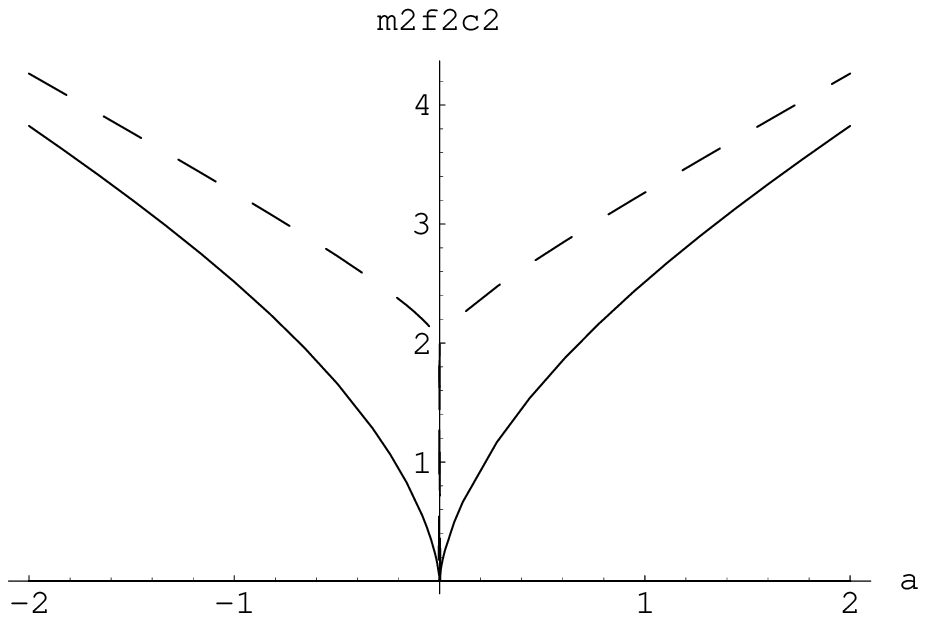}}

\subfigure[Global minimum, $\beta = 3$]{
        \psfrag{A0}[][]{\small $A_m$}
        \psfrag{a}{\small $\alpha$}
        \label{fig:c2lt0:g}
        \includegraphics[width=2.2in]{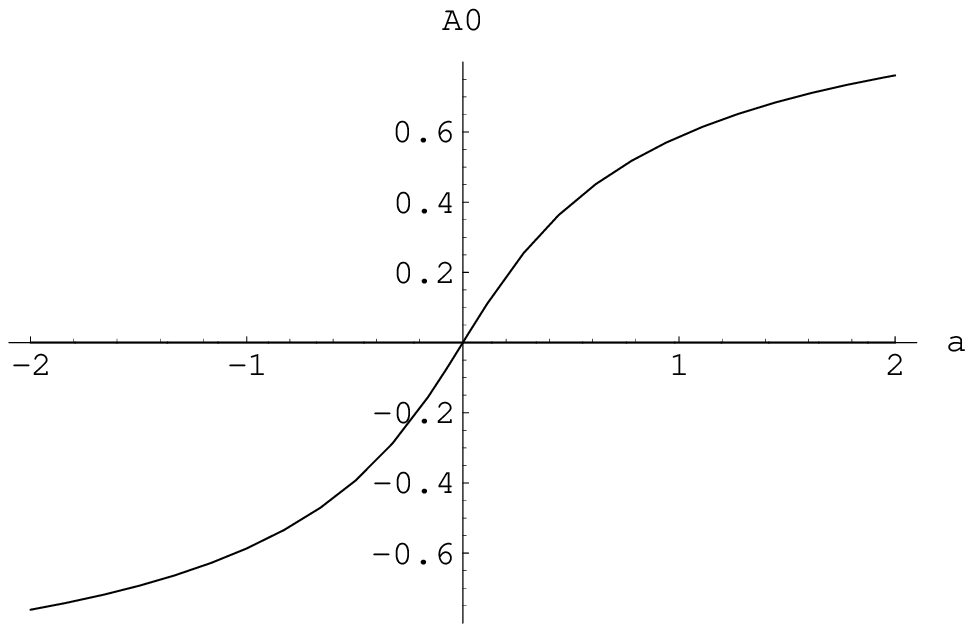}}
\hspace{0.1in}
\subfigure[Pion masses, $\beta = 3$]{
        \psfrag{m2f2c2}[][]{\small $m^2_\pi f^2 / |c_2|$}
        \psfrag{a}{\small $\alpha$}
        \label{fig:c2lt0:h}
        \includegraphics[width=2.2in]{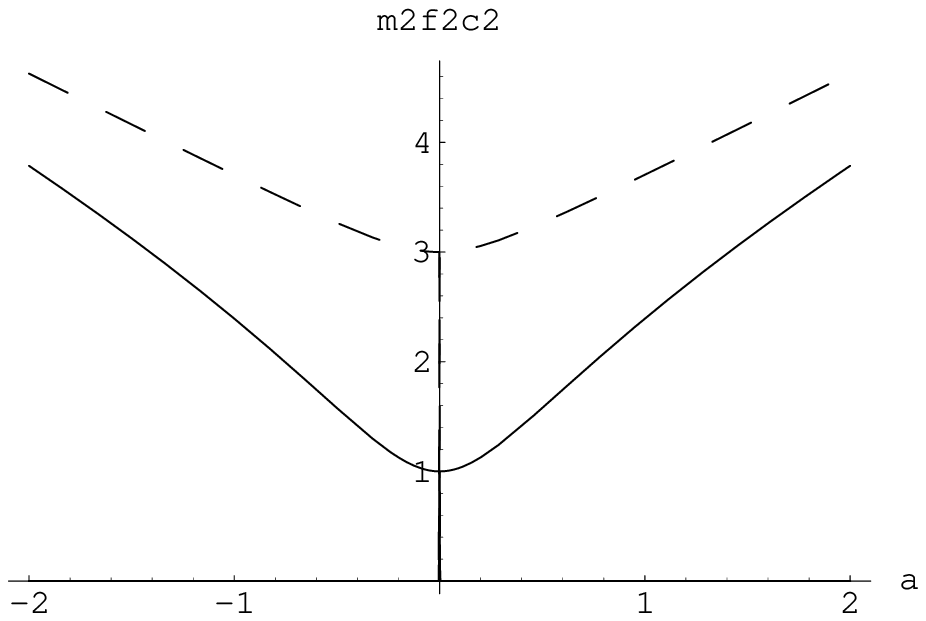}}
\caption{\label{fig:c2lt0} Global minimum $A_m$ and pion masses as a
  function of $\alpha$, for $c_2 < 0$ and $\beta = 0,\,1,\,2,\,3$. The
  dashed lines are for $\pi_{1,2}$ and the solid lines are for $\pi_3$.}
\end{figure}

We now consider the case where $c_2$ is negative.
Fig.~\ref{fig:c2lt0} shows $A_m$ and the pion masses as
a function of $\alpha$ for fixed values of $\beta$.
Fig.~\ref{fig:c2lt0:a} and Fig.~\ref{fig:c2lt0:b} show the results 
for $\mu=0$. As discussed in Ref.~\cite{SS98}, the condensate
jumps from $\Sigma_0 = 1$ (and thus $A_m=1$, $\mathbf{B}=0$) for $\alpha > 0$ 
to $\Sigma_0=-1$ (and thus $A_m = -1$, $\mathbf{B}=0$) for $\alpha < 0$. 
This is a first order transition without flavor breaking, 
so all pions remain massive and degenerate.

The remainder of Fig.~\ref{fig:c2lt0} shows what happens at non-zero
twisted mass. The effect of $\mu$ is to twist the condensate, so that
there is a non-zero $\tau_3$ component $B_{3,m}$.
There is, however, still a first order transition at which $B_{3,m}$
flips sign between $\pm (1-\beta/2)$ (assuming $\beta>0$).
The neutral pion is now lighter than the charged pions due to
the explicit flavor breaking. The neutral pion has a mass
$m_{\pi_3}^2= 2|c_2|(1-\beta/2)^2/f^2$ at the transition, 
while, as noted above, the charged pions have
a constant mass given by $m_{\pi_{1,2}}^2=2 |c_2|/f^2$.
The transition weakens as $|\beta|$ increases, and ends with
a second order transition point at $\beta=\pm 2$,
at which the neutral pion is massless [see Fig.~\ref{fig:c2lt0:f})]. 
For larger $\beta$ the transition is smoothed out.
Note that, once away from the transition, for $\alpha=0$ the
mass-squared splitting between charged and neutral pions is $2 |c_2|/f^2$.

\begin{figure}
\centering
\subfigure[Mass of $\pi_1$ and $\pi_2$]{
\psfrag{m2f2c2}[][]{\large $m^2_\pi f^2 / |c_2|$}
\psfrag{b}{\large $\beta$}
\psfrag{a0}{\small $\alpha = 0$}
\psfrag{a1}{\small $\alpha = 1$}
\psfrag{a2}{\small $\alpha = 2$}
\psfrag{a3}{\small $\alpha = 3$}
\label{fig:c2lt0mpi:a}
\includegraphics[width=3in]{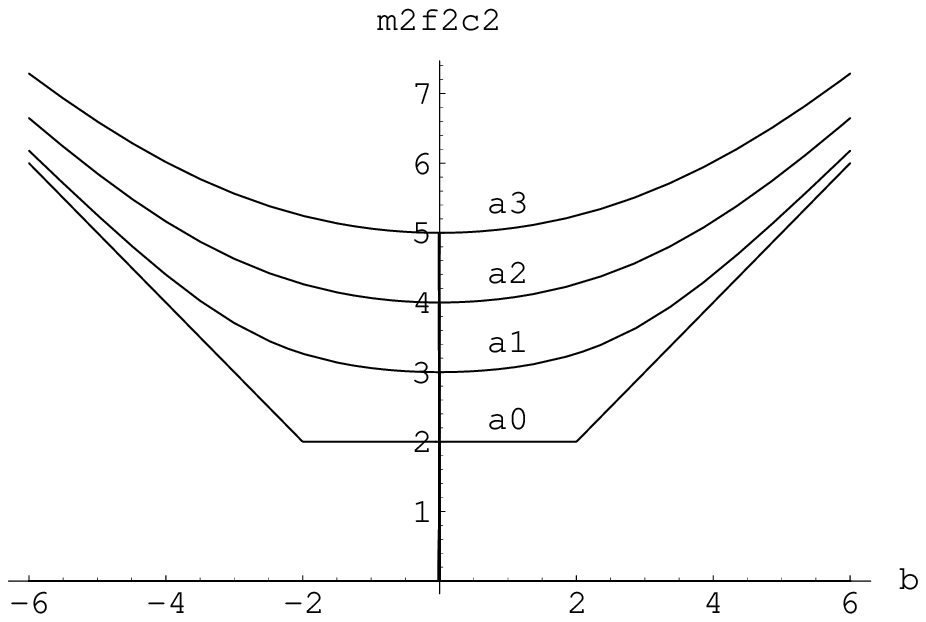}}
\hspace{0.1in}
\subfigure[Mass of $\pi_3$]{
\psfrag{m2f2c2}[][]{\large $m^2_\pi f^2 / |c_2|$}
\psfrag{b}{\large $\beta$}
\psfrag{a0}{\small $\alpha = 0$}
\psfrag{a1}{\small $\alpha = 1$}
\psfrag{a2}{\small $\alpha = 2$}
\psfrag{a3}{\small $\alpha = 3$}
\label{fig:c2lt0mpi:b}
\includegraphics[width=3in]{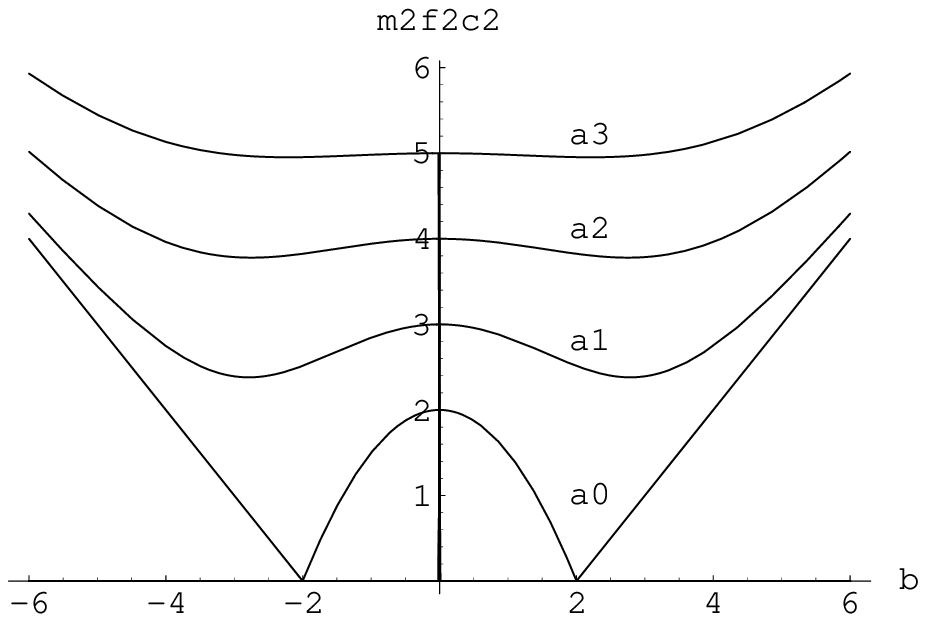}}
\caption{\label{fig:c2lt0mpi} Mass of the pions as a function of
  $\beta$, for $c_2 < 0$ and $\alpha = 0,\,1,\,2,\,3$.} 
\end{figure}

These plots illustrate the general result shown above, namely that
the $c_2<0$ case can be obtained from that with $c_2>0$ by a $90^\circ$
rotation and appropriate redefinitions. Indeed, the results for
$c_2>0$ can alternatively be viewed as the plots for $c_2<0$ at fixed
$\alpha$ with $\beta$ varying, and {\em vice versa}, with the
exception of the charged pion masses, which differ by a constant
offset of $2 |c_2|/f^2$. To illustrate this latter point we
plot, in Fig.~\ref{fig:c2lt0mpi}, the pion masses for $c_2<0$ as a
function of $\beta$ for fixed values of $\alpha$.
Comparing to Fig.~\ref{fig:c2gt0mpi}, we see the equality of
the neutral pion masses and the constant offset in the charged pion masses.

\begin{acknowledgments}
This research was supported in part by the U.S. Department of Energy
Grant No. DE-FG02-96ER40956. We thank Oliver B\"ar,
Roberto Frezzotti and Gernot M\"unster
for useful conversations.
\end{acknowledgments}

\bibliography{paper}

\end{document}